\documentclass[runningheads,a4paper]{llncs}

\usepackage{graphicx}
\usepackage{todonotes}
\usepackage{amsmath}
\usepackage{amssymb}

\graphicspath{{img/}} 

\newcommand{\keywords}[1]{\par\addvspace\baselineskip
\noindent\keywordname\enspace\ignorespaces#1}

\begin{document}
\mainmatter 

\title{Cheetah Experimental Platform Web 1.0: Cleaning Pupillary Data}
\titlerunning{Cheetah Experimental Platform Web 1.0}
\authorrunning{S. Zugal \and J. Pinggera  \and M. Neurauter \and T. Maran \and Barbara Weber}
\author{Stefan Zugal\inst{1}  \and Jakob Pinggera\inst{1}  \and Manuel
Neurauter\inst{1} \and Thomas Maran\inst{1} \and Barbara Weber\inst{1,2}}

\institute{
	University of Innsbruck, Austria\\
\email{Firstname.Lastname@uibk.ac.at}
\and
Technical University of Denmark, Denmark\\
\email{bweb@dtu.dk}}	
\maketitle

\begin{abstract}
Recently, researchers started using cognitive load in various settings, e.g., educational psychology, cognitive load theory, or human--computer interaction. Cognitive load characterizes a tasks' demand on the limited information processing capacity of the brain. The widespread adoption of eye--tracking devices led to increased attention for objectively measuring cognitive load via pupil dilation. However, this approach requires a standardized data processing routine to reliably measure cognitive load. This technical report presents CEP--Web, an open source platform to providing state of the art data processing routines for cleaning pupillary data combined with a graphical user interface, enabling the management of studies and subjects. Future developments will include the support for analyzing the cleaned data as well as support for Task--Evoked Pupillary Response (TEPR) studies.

\end{abstract} 

\section{Introduction}

In recent years, research started to focus on cognitive load in various settings such as in educational psychology \cite{Paas2003}, cognitive load theory~\cite{Sweller2011}, or human computer interaction \cite{Bailey2008}. In general, cognitive load characterizes the demands of tasks imposed on the limited information processing capacity of the brain \cite{Wickens2012}. Cognitive load therefore represents an individual measure depending on available resources interacting with given tasks on a subjective level. Four main approaches to measure cognitive load are common in research. For a detailed overview of cognitive load measures see \cite{Chen2016}. 

\begin{description}
\item[Subjective measures]\hfill \\
Subjects are ranking their experienced level of cognitive load using rating scales.
\item[Performance measures]\hfill \\ 
For example task performance of secondary tasks, critical errors, task completion times, or speed.
\item[Physiological measures]\hfill \\ 
Such as heart rate or heart rate variability, galvanic skin response, or pupil dilation.
\item[Behavioural measures]\hfill \\ 
Observing behavioural patterns such as mouse-click events or linguistic patterns.
\end{description}

The easier access to eye-tracking devices led to increasing attention for measuring cognitive load via pupil dilation in recent years \cite{Chen2016}. In general, pupillary responses can have a variety of causes such as reflex reaction based on light exposure \cite{Ellis1981} or sexual stimulation \cite{Hess1960}. Additionally, pupil dilation also is caused by cognitive load as shown by Beatty \cite{Beatty1982}, who identified pupil dilation to be a reliable indicator of processing load during tasks. Flowing this insights, several studies utilized pupil dilation measurements to assess cognitive load in various settings (e.g. \cite{Bailey2008,Hayes2016,Koelewijn2014}). 

Pupil dilation data however, needs to run through several cleaning and filtering routines to reliably measure cognitive load \cite{Ped+11}. To the best of our knowledge, no common data cleaning and filtering software package is available, resulting in a variety of different techniques and levels of preprocessing pupillary data. Therefore, this technical report describes a state of the art preprocessing routine for pupillary data. To support the execution of the described preprocessing, we introduce Cheetah Experimental Platform Web 1.0 (CEP--Web), following and enhancing the work presented in~\cite{Weber2015,Neurauter2015}.  

Following a generic approach, eye-tracking data in stimulus response setting common in psychology as well as data gathered in long running tasks can be prepared for following analyses using CEP--Web. Due to the high amounts of data gathered by eye-tracking systems working at high sampling rates (e.g. 300 Hz for the Tobii TX300) the tool is installed on a server at the University of Innsbruck and therefore able to perform the needed calculations without imposing load on researchers IT components. Interested researcher might also obtain the source code of CEP--Web via our GitLab page\footnote{https://git.uibk.ac.at/cheetah-web-group/cheetah-web}. Along with the possibility of calculating the cognitive load within data sets, the opportunity to visualize both, the raw data and the processed data is provided. The visualization of the data allows to verify the changes during the preprocessing and offers a possibility to examine the user's cognitive load trend directly. 

We hope providing this infrastructure is aiding the usage of pupil dilation data to assess cognitive load in various fields of research by lowering the demands of needed computing performance and programming skills for researchers. This should lead to more research utilizing cognitive load in various scientific fields, providing deeper insights into human-computer interaction, software design, or design research.  

In the remainder of the report is structured as follows. Section~\ref{sec:analysis-workflow} describes the CEP--Web's analysis workflow and the available pupillometric cleaning routines. Section~\ref{sec:technical-infrastructure} presents the  technical infrastructure of CEP--Web. Section~\ref{sec:related_work} outlines related work and Section~\ref{sec:summary} concludes the technical report with a brief summary and outlook on future extensions of CEP--Web. 

\section{Analysis workflow} \label{sec:analysis-workflow}
\begin{figure}[htp]
\begin{center}
  \includegraphics[width=0.8\textwidth]{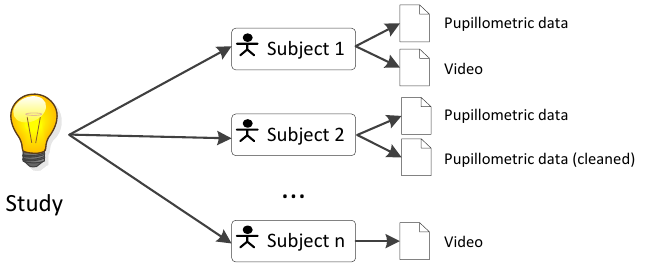}
  \caption{Study with subjects and files}
  \label{fig:design_overview}
\end{center}
\end{figure}

This section describes the workflow and corresponding features for analyzing
pupillometric data. Particularly, experimental data is
organized in CEP--Web as illustrated in Figure~\ref{fig:design_overview}: the
for each study to be analyzed using CEP--Web, one or more subjects may be assigned.
To each subject, in turn, multiple files can be attached---ranging from
\textit{raw} input data exported from the eye tracker and associated video data
as well as data that was already processed in CEP--Web.

Based on this structure, we recommend the following workflow for the analysis of
pupillometric data within CEP--Web:

\begin{itemize}
  \item \textit{Create study and subject:} As first step, we recommend to create
  a study for the experiment to be analyzed and also to create all subjects that
  participated in the study. 
  \item \textit{Upload raw data:} In the second step, export the data from the
  eye tracker and upload it to CEP--Web. To help with the organization of files,
  it is mandatory to assign each file to a subject; please note that a subject
  may have multiple files assigned.
  \item \textit{Cleaning:} After all data is uploaded, CEP--Web provides a
  number or filters for processing and cleaning pupillometric data.
  \item \textit{Data inspection:} To visually inspect whether the cleaning
  achieved the desired results, CEP--Web provides the possibility to inspect raw
  and cleaned data by visualizing the data as graphs.
  \item \textit{Pupil size calculation:} CEP--Web 1.0 provides an initial,
  rudimentary implementation for computing the average pupil sizes. 
\end{itemize}

In the following, we detail each of the steps. Particularly,
Section~\ref{sec:subject_management} covers study and subject management,
Section~\ref{sec:file_management} explains the management of data files, whereas
Section~\ref{sec:pupillometric_cleaning} details the cleaning of pupillometric
data. Finally, Section~\ref{sec:data_inspection} illustrates how CEP--Web
supports the visual inspection of raw and cleaned data.

\subsection{Study and subject management}
\label{sec:subject_management}
The management of subjects is the first step for analyzing data in CEP--Web.
In particular, for each experiment that should be analyzed, we recommend to
create an individual study and assign all subjects that participated in the
study. To facilitate the creation of subjects, an import feature was implemented allowing the creation of several subjects by importing them using a comma--separated values (csv) file. 

\subsection{Data management}
\label{sec:file_management}
The file management of CEP--Web allows for uploading and administering files.
Even though the naming of files is up to the user, for the purpose of
maintainability, we recommend to follow a certain pattern for naming files:

\begin{center}
\noindent\texttt{\centering{subject\_id@study.extension}}
\end{center}

Thereby, \textit{subject\_id} refers to the identifier of the subject, e.g.,
subject number, \textit{study} refers to the study the subject participating and
\textit{extension} to the extension of the file. In this sense,
pupillometric data of subject number \textit{16} participating in study
\textit{modeling\_experiment} should be named as follows:

\begin{center}
\noindent\texttt{{16@modeling\_experiment.tsv}}
\end{center}

By adhering to this naming convention, the files will be automatically mapped to the previously created subjects. Otherwise, the mapping of file to subject has to be established manually. 

\subsection{Pupillometric cleaning}
\label{sec:pupillometric_cleaning}
For the cleaning of pupillometric data, which is the centerpiece of CEP--Web
1.0, we have implemented a series of \textit{filters}. Basically, a filter
takes pupillometric data---left pupil size, right pupil size and timestamp---as
input and applies a set of transformation steps on this data. The output provided by one
filter can then be directly fed into the next filter, allowing for establishing
a filtering chain. Being able to chain filters is of particular interest, since
each filter serves a particular purpose in the process of cleaning pupillometric
data. Certain filters remove data points, e.g., by removing
outliers, whereas other filters are designed to interpolate missing values. In
other words, typically filters removing data are applied before filters that
interpolate data, since generally the output of the cleaning process should be a
continuous series of data without missing values.

Even though CEP--Web allows for freely combining filters, it should be
emphasized that not all combination of filters will lead to satisfying results.
For instance, applying the Butterworth filter to data that includes missing
values, will lead to filtering artifacts. Contrariwise, the blink detection
filter will not work properly if missing values---which are used for identifying
blinks---have been linearly interpolated before. Therefore, we recommend to
apply filters in the following order:

\begin{itemize}
  \item \textit{Pupil substitution (optional)}: Basic interpolation filter for
  pupil sizes.
  \item \textit{Gazepoint substitution (optional)}: Basic interpolation filter
  for gaze points.
  \item \textit{Blink detection}: Detect and remove blink artifacts.
  \item \textit{Standard deviation}: Remove outliers identified by statistics.
  \item \textit{Linear interpolation}: Linearly interpolate missing values.
  \item \textit{Butterworth Filter}: Applies a third order lowpass Butterworth filter
\end{itemize}

For most purposes, the ordering described above will lead to the
best results. However, in certain scenarios, it may be useful to shift or omit
specific filters. To understand, which filters can be applied for which purpose,
we describe all filters provided by CEP--Web in the following.

\subsubsection{Pupil substitution.}
The pupil substitution filter is one of the basic filters for interpolating
missing data and can be applied for binocular eye trackers only. The idea behind
this filter is to substitute missing values from one pupil by the value measured
for the other pupil. Hence, the left pupil size is substituted with
the right pupil size, of value of the left pupil size is missing and the value of
the right pupil size is not missing:

\begin{equation}
filter(value_\textit{left}) = 
\begin{cases}
value_\textit{right} & \text{if } missing(value_\textit{left})\ \wedge \
!missing(value_\textit{right})\\
value_\textit{left} & \text{else}
\end{cases}
\end{equation}

Analogous, the right pupil size is replaced with the left pupil size, if the
value of the right pupil size is missing and the value of the left pupil size is
not missing:

\begin{equation}
filter(value_\textit{right}) = 
\begin{cases}
value_\textit{left} & \text{if } missing(value_\textit{right})\\
value_\textit{right} & \text{else}
\end{cases}
\end{equation}

We would like to mention at this point that the substitution filter should be
applied with care, since the size of the left and right pupil may differ
considerably.

\subsubsection{Gazepoint substitution.}
Similar to the pupil substitution filter, the gaze point substitution filter will
replace gaze points, if possible. In particular, we consider a gaze point
with screen coordinates (x, y) missing if either the value of x (horizontal
position in pixel) or y (vertical position in pixel) is missing:

\begin{equation}
missing(x, y) = 
\begin{cases}
true & \text{if } x\ missing \vee y \ missing \\
false & \text{else}
\end{cases}
\end{equation}

Based on this definition of missing, we substitute the left gaze point with the
right gaze point, if the value for the left gaze point is missing and the value
for the right gaze point is not missing:

\begin{equation}
filter(gaze_\textit{left}) = 
\begin{cases}
gaze_\textit{right} & \text{if } missing(gaze_\textit{left})\ \wedge\ 
!missing(gaze_\textit{right})\\
gaze_\textit{left} & \text{else}
\end{cases}
\end{equation}

Analogous, we substitute the right gaze point with the
left gaze point, if the right gaze point is missing and the left gaze point is
not missing:

\begin{equation}
filter(gaze_\textit{right}) = 
\begin{cases}
gaze_\textit{left} & \text{if } missing(gaze_\textit{right})\ \wedge\ 
!missing(gaze_\textit{left})\\
gaze_\textit{right} & \text{else}
\end{cases}
\end{equation}

\subsubsection{Standard deviation.}
The assumption behind the standard deviation filter is that all values that
differ more than 3 times from the mean value, should be considered outliers and
are removed. More formally:

\begin{equation}
filter(value) = 
\begin{cases}
missing & \text{for } value > \bar{x} + 3 * \sigma \\
missing & \text{for } value < \bar{x} - 3 * \sigma \\ 
value & \text{else}
\end{cases}
\end{equation}

\begin{figure}
\centering
\begin{minipage}{0.45\textwidth}
\centering
\includegraphics[width=\textwidth]{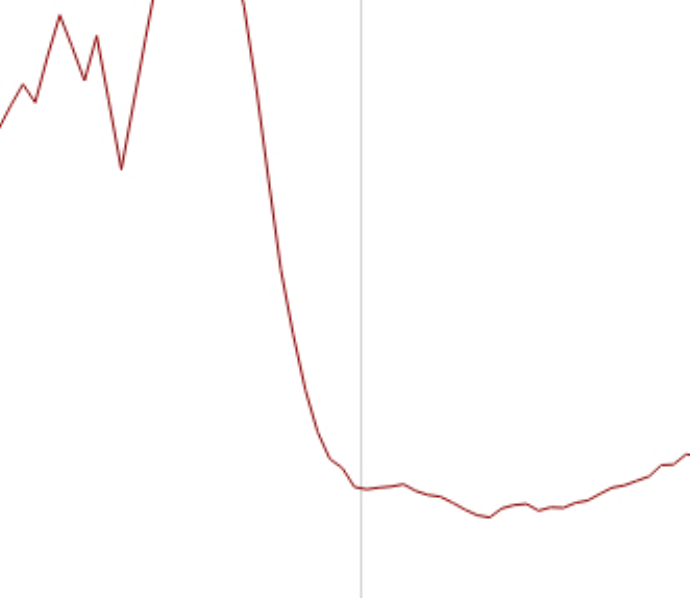}
\caption{Raw data}
\label{fig:standard_deviation_raw}
\end{minipage}\hfill
\begin{minipage}{0.45\textwidth}
\centering
\includegraphics[width=\textwidth]{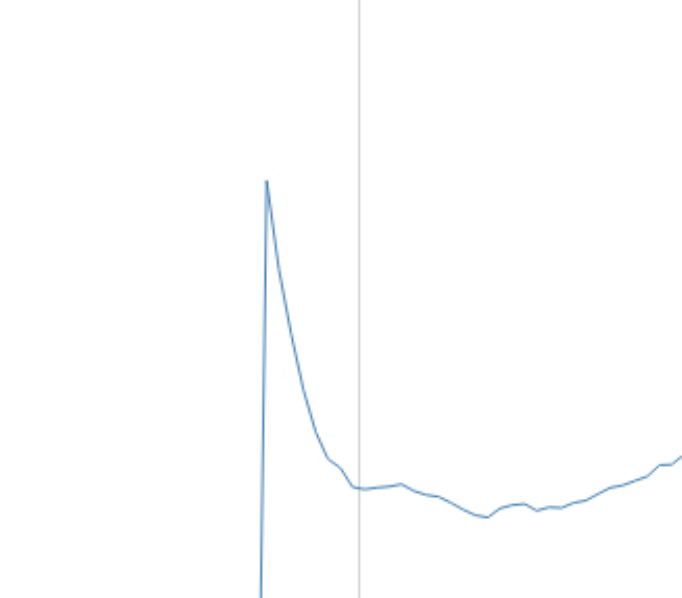}
\caption{After applying standard deviation filter}
\label{fig:standard_deviation_filtered}
\end{minipage}
\end{figure}

To illustrate the application of the standard deviation filter, consider the
graphs in Figures~\ref{fig:standard_deviation_raw} and
\ref{fig:standard_deviation_filtered}. Particularly, the graph in
Figure~\ref{fig:standard_deviation_raw} shows a distortion to an extent that the
visualization engine of CEP--Web could not print the graph anymore (all filter
examples shown in this document are screenshots from CEP--Web). After applying
the standard deviation filter, these anomalies are removed---the resulting
missing values could then be substituted by a linear interpolation filter in a
second step.

\subsubsection{Blink detection.}
\begin{figure}
\centering
\begin{minipage}{0.45\textwidth}
\centering
\includegraphics[width=\textwidth]{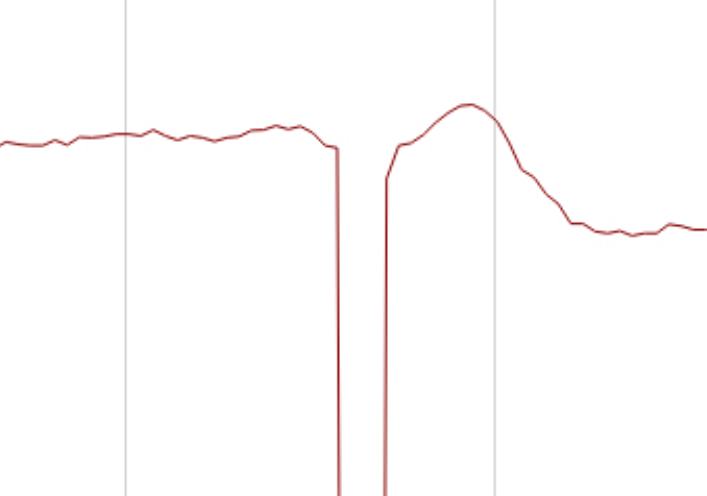}
\caption{Raw data}
\label{fig:blink_raw}
\end{minipage}\hfill
\begin{minipage}{0.45\textwidth}
\centering
\includegraphics[width=\textwidth]{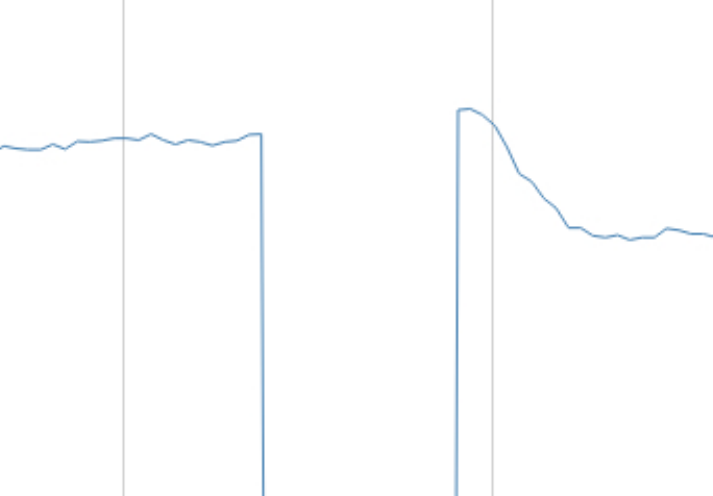}
\caption{After applying blink detection}
\label{fig:blink_filtered}
\end{minipage}
\end{figure}

The detection of blinks is one of the central aspects when cleaning
pupillometric data. Since the blink itself is not an on--off event, but rather a
process of closing and opening the eyelid, artifacts are likely to occur shortly
before and after the blink. To compensate for these artifacts, CEP--Web
implements the blink detection filter, which, based on a
heuristic of missing values and gaze position, will detect and clip
out blinks (for details, please refer to~\cite{Ped+11}). To illustrate the
application of the blink detection filter, consider the
Figures~\ref{fig:blink_raw} and \ref{fig:blink_filtered}. As indicated before,
the filter recognizes the series of missing data as blink and clips out all
values that were measured during blink onset and blink ending.

In case the pupil substitution filter or the gaze point substitution filter is
intended to be applied before the blink detection filter, we \textit{strongly
recommend} to apply these filters only in combination. As described
in~\cite{Ped+11}, for detecting blinks, the blink detection filter takes into
account whether pupil sizes are available and whether a valid gaze point
position could be measured. Hence, the filter will only work properly, if both
pupil sizes and gaze points have been substituted. To clarify, the application
of pupil substitution filter or gaze point substitution filter is \textit{not}
mandatory, but they should only be applied in combination when the blink
detection filter should be applied as well.

\subsubsection{Linear interpolation.}
Whenever eye trackers are used for collecting pupillometric data, data loss is
unavoidable---simply due to the fact that the eye tracker cannot assess the
pupil during blinks. This is particularly problematic since certain filters
depend on continuous data, such as the the Butterworth filter. To deal with this
situation, CEP--Web provides a filter for linearly interpolating missing data.
Particularly, for a series of missing values, where $value_{s}$ is the
last non-missing value at time $time_{s}$ and
$value_{e}$ is the first non-missing value after a series of missing values
with the corresponding timestamp $time_{e}$, CEP--Web will linearly
interpolate the missing values as follows:

\begin{equation}
value(time) = value_{s} + (value_{e} - value_{s}) * \frac{time -
time_{s}}{time_{e} - time_{s}}
\end{equation}

\begin{figure}
\centering
\begin{minipage}{0.45\textwidth}
\centering
\includegraphics[width=\textwidth]{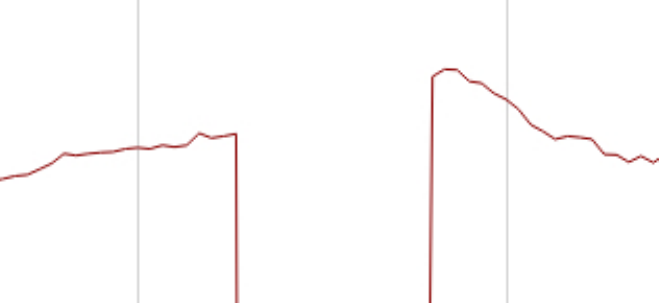}
\caption{Raw data}
\label{fig:linear_interpolation_raw}
\end{minipage}\hfill
\begin{minipage}{0.45\textwidth}
\centering
\includegraphics[width=\textwidth]{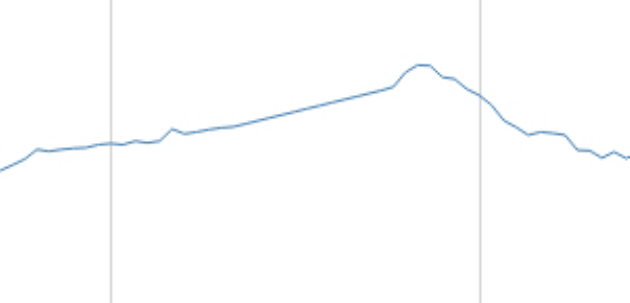}
\caption{Linearly interpolated data}
\label{fig:linear_interpolation_filtered}
\end{minipage}
\end{figure}

To illustrate linear interpolation in CEP--Web, consider
Figures~\ref{fig:linear_interpolation_raw} and
\ref{fig:linear_interpolation_filtered}. Clearly,
Figure~\ref{fig:linear_interpolation_raw} is affected by missing data---in this
particular example caused by a blink. After applying the filter, the missing
data is linearly interpolated, resulting in continuous data, as shown in
Figure~\ref{fig:linear_interpolation_filtered}.

When applying the linear interpolation filter, we \textit{strongly recommend} to
apply the blink detection filter first. In particular, values measured during
blinks are error--prone due the eyelid being halfway covered, hence the linear
interpolation filter should not use these values as start/end points for linear
interpolation.
Rather, values that can be considered valid, i.e., after the pupil is fully
visible and the eye tracker can assess the pupil size correctly, should be used
as start/end points for linear interpolation. For achieving this, the blink
detection filter can be used, since it will remove values that were measured
during a blink, hence removing values that were measured when the eyelid was
closing/opening.

Similarly, the eye tracker may deliver spurious measurements during blinks,
i.e., the eye tracker incorrectly detects the pupil. Again, respective values
must be assumed to be invalid and should not be used for linear
interpolation---also for this reason, we \textit{strongly recommend} to apply
the blink detection filter before running the linear interpolation.

\subsubsection{Butterworth filter.}
\begin{figure}
\centering
\begin{minipage}{0.45\textwidth}
\centering
\includegraphics[width=\textwidth]{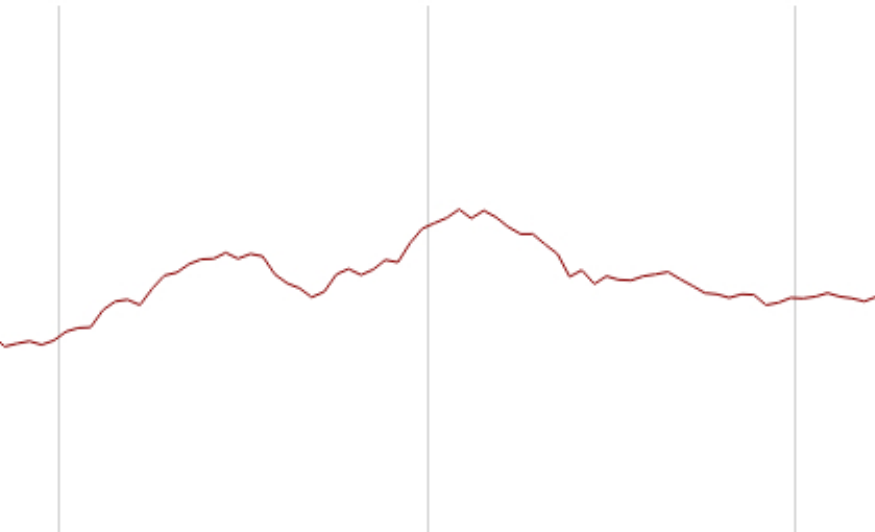}
\caption{Raw data}
\label{fig:butterworth_raw}
\end{minipage}\hfill
\begin{minipage}{0.45\textwidth}
\centering
\includegraphics[width=\textwidth]{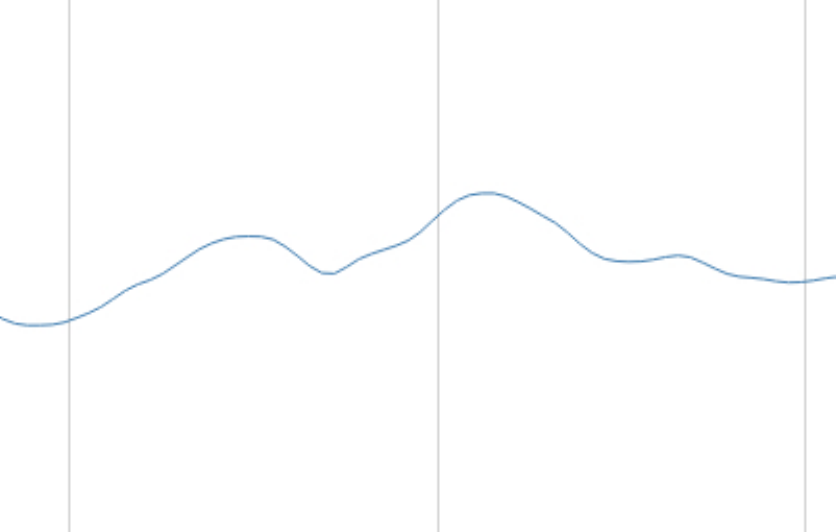}
\caption{Filtered with Butterworth}
\label{fig:butterworth_filtered}
\end{minipage}
\end{figure}

CEP--Web provides a third order lowpass Butterworth filter for data cleaning
through an open--source library for digital signal
processing\footnote{\url{http://www.source-code.biz/dsp/java}}. Generally,
lowpass filters will filter out high frequencies, but do not affect lower
frequencies. For data cleaning, lowpass filters are typically used to filter out
measurement artifacts such as white noise and are thus used for smoothening
data. To illustrate the effect of applying a lowpass filter, consider the
illustrations shown in Figures~\ref{fig:butterworth_raw} and
\ref{fig:butterworth_filtered}.
The raw data shows typical signs of white noise, whereas the filtered data
exhibits the same basic features, but is apparently smoothened.

Even though a lowpass Butterworth filter can be used for processing measurement
artifacts, its application introduces a phase response to the filtered
signal---in other words, this feature of Butterworth causes the filtered signal to be
shifted towards the past. To compensate for this phase response, CEP--Web
calculates the expected phase response and automatically re--shifts the
processed signal so that the phase response is equalized. In particular, the
phase shift of a third order lowpass Butterworth filter ($\phi$) can be
calculated based upon the angular frequency ($w$) as follows:

\begin{equation}
\phi(w) = - tan^{-1}\Bigg(\cfrac{2w - w^3}{1 - 2w^2}\Bigg)
\end{equation}

\subsection{Data inspection}
\label{sec:data_inspection}
\begin{figure}[htp]
\begin{center}
  \includegraphics[width=\textwidth]{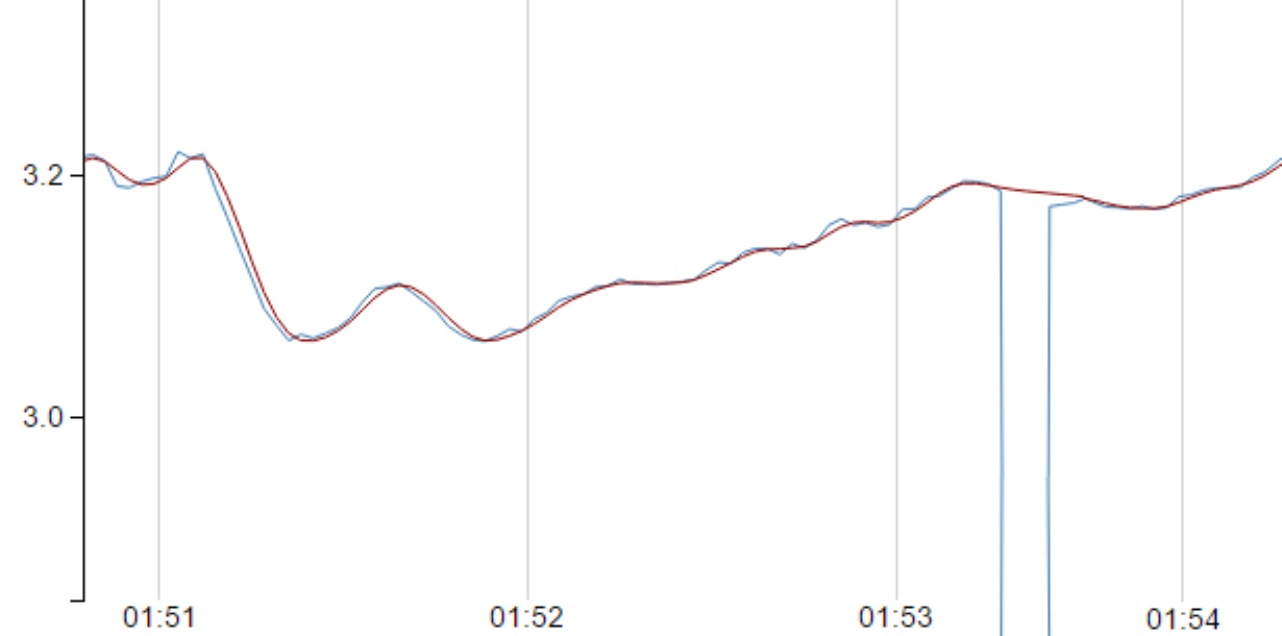}
  \caption{Visualization of raw data (blue) and cleaned data (red)}
  \label{fig:visualization}
\end{center}
\end{figure}

As described in Section~\ref{sec:pupillometric_cleaning}, filters need to be
applied with care and the effect of applying a filter is not always obvious.
Similarly, it is difficult to assess whether the data being analyzed behaves the
way it was expected for the experiment or whether it contains unforeseeable
artifacts. To counteract this problem, CEP--Web provides the possibility of
visualizing pupillometric data in the form of graphs, as illustrated in
Figure~\ref{fig:visualization}. In this particular example, the user chose to
visualize raw data (blue line) to compare them with the cleaned results (red
line); on the x-axis the duration is shown in minutes:seconds, on the y-axis the
pupil diameter is shown in mm. Furthermore, it can be seen that the cleaning
resulted in a smoothening of the graph around 1:51 and the removal of a blink
around 1:54.

\section{Technical infrastructure} \label{sec:technical-infrastructure}
This section explains on the technical infrastructure provided by CEP--Web.
Section \ref{sec:technical-challenges} focuses on the technical challenges
connected with processing large amounts of data and describes in Section
\ref{sec:software-architecture} how CEP--Web tackles these challenges and
supports the analysis workflow described in Section \ref{sec:analysis-workflow}.

\subsection{Technical Challenges}
\label{sec:technical-challenges}
Besides typical challenges of software, such as maintainability, quality and
usability, CEP--Web needs to address two challenges particular to the processing
of pupillometric data:

\subsubsection{Data volume.} Obviously, the amount of data to be processed  
strongly depends on the extend of data provided by the eye tracker. This, in  
turn, is typically mainly influenced by the sampling rate, data fields  
exported from the eye tracker, as well as the format used for export. First,  
regarding the sampling rate, modern eye trackers work with a sampling rate of  
at least 60Hz, i.e., producing 60 data points per second.
Depending on the particular   field of application, sampling rates of up to
1000Hz are common as well. CEP--Web does not make any assumptions about sampling
rates and thus supports arbitrary sampling rates. Second, regarding the data
fields exported by the eye tracker, it mainly depends on the eye tracking
software provided with the eye tracker. Even though the export may be configured
to that only necessary fields will be exported by the eye tracking software, for
convenience or simplicity often all available data is exported. Here, the data
fields required by CEP--Web depend on the particular application. For instance,
for rather simple filters such as standard deviation, timestamps as well as the
pupillometric data are sufficient. However, for more complex filters, such as
the blink detection, additionally gaze positions may be required. CEP--Web does
not require the user to specify theses data fields up--front, but will prompt
the user as soon as they are required. Third, regarding the data format,
compressed or proprietary formats may provide a more succinct representation of
data. Since compressed or proprietary formats tend to compromise
interoperability, in CEP--Web, we rely on tabular separated value (.tsv) files
only.

Given our typical usage scenario with Tobii X300 and considering a sampling rate
of 300Hz, one hour of eye tracking will produce about 430MB of data. In the
light of our recent study with 115 subjects and eye tracking sessions between 30
minutes and more than 1 hour, 36.34 GB of data needed be processed. Given these
numbers, it becomes evident that more elaborated approaches are required to
handle this amount of data.

\subsubsection{Processing speed.} Closely connected to the volume of data is the
time and processing power required for analyzing respective data. Apparently,
the more data needs to be processed and the more complex the computations to be
conducted, the longer the data analysis will take. Given the amount of data
at hand, optimizing the individual computations (e.g., cleaning or filtering) is
not enough to keep the processing duration at an acceptable level. Rather, it is
of vital importance to provide support for the parallel processing of
computation for providing reasonable computation times.

To exemplify the need for parallel processing, consider our recent study in
which the have tracked 115 subjects with a sampling rate of 300Hz, resulting in
a data set with 96,676,561 data points. For each of these data points, we were
required to run the pupil substitution filter, standard deviation filter, linear
interpolation filter as well as Butterworth filter. Apparently, running this
kind of analysis on a single CPU will not be enough to provide results in 
acceptable time.

\subsection{Software architecture}
\label{sec:software-architecture}
To address the described challenges with respect to data volume and processing
speed, we have developed CEP--Web that allows for data inspection in realtime,
whenever possible, but also provides the possibility to schedule long--lasting
tasks in the background. Particularly, as described in the following, we provide
a worker infrastructure for parallelizing long--lasting tasks, as well as means
for pre--processing data for real--time analysis.

\subsubsection{Worker infrastructure.}
\begin{figure}[htp]
\begin{center}
  \includegraphics[width=\textwidth]{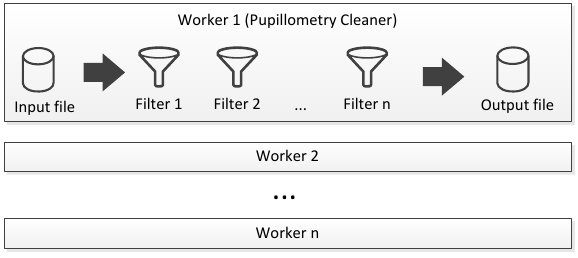}
  \caption{Worker infrastructure}
  \label{fig:worker}
\end{center}
\end{figure}

Even though the clock speed of modern CPUs is currently restricted by physical
limits and cannot be expected to be increased any further, modern computers
provide performance gains through the parallel operation of several CPU cores.
However, to harness the power of parallel processing, the software needs to
provide the respective infrastructure as well. In CEP--Web, we employ the thread
pool pattern (also known as replicated workers or worker--crew
model)~\cite{GaSh02} to parallelize the long--lasting computations. In
particular, we seek to encapsulate each task in the form of an independent
\textit{worker}, as illustrated in Figure~\ref{fig:worker}. Therein, the task of
\textit{Worker 1} is to read an input file, apply a set of \textit{n} filters to
the input file and finally write the processed file as output of the worker. In
parallel to \textit{Worker 1}, several other workers may be processing data as
well---again, each worker will encapsulate the work to be done and can be run as
independent computation. 

The crucial part of this worker infrastructure is the scheduling of work to be
conducted. Too few parallel workers will result in a low utilization of CPUs,
while a too high number of parallel workers will negatively impact performance
when several workers need to share on CPU. In CEP--Web we decided to allow
exactly one worker on single--core machines. On multi--core machines, we reserve
one core for housekeeping work and use the remaining cores for workers, i.e.,
the number of workers in parallel is the number number of CPU cores - 1. Workers
that cannot be started yet due to this restriction are kept in the worker queue
and will be started as soon as one of the active workers has finished.

\subsubsection{Preparation of data for real--time analysis.}
As described in Section~\ref{sec:analysis-workflow}, CEP--Web also provides the
possibility to inspect raw as well as processed data in real--time. As described
in Section~\ref{sec:technical-challenges}, particularly eye trackers with high
sampling rates easily produces data files are several hundreds of MB large. Even
though modern hard drives can easily store and read this amount of data, it
still requires the user to wait until the system has loaded and processed the
file. To minimize this waiting time and enable real--time data inspection, data
files are optimized along three stages:

\begin{itemize}
  \item \textit{Uncompressed data.} Data that is uploaded by the user or being
  stored from a worker as result of applying filters typically contains data
  fields that are not necessary for real--time data inspection and implies that
  unnecessary data is loaded from the hard disk. For performance reasons,
  CEP--Web does not provide real--time data inspection for these files.
  \item \textit{Compressed data.} To make uncompressed data amenable for
  analysis, CEP--Web  strips of all unnecessary data from uncompressed files
  and stores these compressed files in new, \textit{compressed} data files.
  Since this typically involves reading, processing and storing several hundreds
  of MBs, the compressing of data is executed as a task within the worker
  infrastructure.
  \item \textit{Cached data.} Even though compressed data can be read faster
  than uncompressed data, the waiting times for disk I/O operations are still to
  large for ensuring acceptable waiting times during real--time analysis. To
  finally achieve acceptable waiting times, CEP--Web holds the data currently
  under analysis in memory to allow for instant access. Hence, whenever a data file
  should be analyzed, its content is loaded into memory, imposing a onetime
  waiting time. For the remainder of the analysis, data will be read from
  memory, thereby miniming waiting times.
\end{itemize}

\subsection{Implementation}
\label{sec:implementation}
CEP--Web is implemented as typical web application on top of Tomcat on the
server side and relying on AngularJS on the client side. In the following, we
will describe details that may be interesting for readers interested in extended
CEP--Web. For a more up--to--date technical description of CEP--Web, please
visit our GitLab
page\footnote{https://git.uibk.ac.at/cheetah-web-group/cheetah-web}.

\subsubsection{Server side.} CEP--Web builds upon Tomcat 6 and Java 7. For
storing data, we rely on MySQL 5.6; access to the database is provided by the
MySQL JDBC driver. For migrating our database, we rely on LiquiBase;
respective changelogs are available in our GitLab account. CEP--Web does not
have any operating system depending components and has so far been used on
Windows 10 and CentOS 6.7.

\subsubsection{Client side.} For the client side, we rely on AngularJS and
Bootstrap. Hence, CEP--Web generally supports all our modern browsers---best
results can be expected with Google Chrome and Mozilla Firefox.

\subsection{Hardware setup}
\label{sec:hardware_setup}
The minimum requirements on hardware are difficult to describe, since they
heavily depend on the data to be analyzed and the filters that should be
applied. From our experience on analyzing data, we know:
\begin{itemize}
  \item The need of memory for cleaning data in parallel workers does not
  increase linearly with each worker. Rather, the first worker will require a
  proportionally larger share of memory than any additional worker. Technically,
  this behaviour can be explained by the effect of sharing String objects
  (String pooling) that is performed by Java, i.e., identical Strings are not
  recreated, but rather shared throughout the virtual machine. Hence, the first
  worker will fill the String pool---requiring proportionally more
  memory---whereas any additional worker will reuse the Strings from the pool.
  \item For our recent study, we cleaned data with up to 500MB on a machine with
  24 GB RAM and 12 cores. As described in
  Section~\ref{sec:software-architecture}, on this setup, only 11 cores will be
  used at the same time for cleaning, i.e., at most 11 workers will run in
  parallel. For this amount of workers, the memory was easily sufficient for
  cleaning the data and computing the average pupil size.
\end{itemize}
\section{Related work}
\label{sec:related_work}

In \cite{Koelewijn2014} several preprocessing steps are described in a generic way, such as blink detection, linear interpolation in blink phases, and removing high frequency artifacts. All of those are addressed by CEP--Web as described in Section~\ref{sec:analysis-workflow}. The eye blink detection approach presented in \cite{Ped+11} and a corresponding handling of blinks during eye tracking is implemented in CEP--Web. To remove high frequency artifacts we are following \cite{Jiang2015} by using a third order lowpass Butterworth filter also described in Section~\ref{sec:analysis-workflow}.

For an exhaustive overview regarding the theory for cognitive load, it's measurement and examples for research please see \cite{Chen2016}.
\section{Summary}\label{sec:summary}
This technical report provided a brief overview of the current capabilities of CEP--Web in terms of managing, cleaning, and visualizing pupillometric data. More specifically, we presented the handling of studies and subjects and their associated pupillometric data. Further, the available filters of CEP--Web are presented. This way, users are capable of performing state of the art data processing, even on large numbers of files. The integrated visualization tool allows users to inspect their data prior and after cleaning to ensure the correct application of the filters.

Future developments for CEP--Web will include the evaluation of pupil dilation during tasks and sub--tasks. For this, we are planning an infrastructure that splits the data stream into smaller chunks and provides capabilities for calculating and reporting mean, median, and/or standard deviation. In a similar vein, we intend to include support for facilitating the analysis of Task--Evoked Pupillary Response (TEPR) studies. For this, we plan to integrate the definition of data processing routines that include not only cleaning, but also partitioning the data into trials and evaluating the pupillary responses after the stimulus. The data will be aggregated and presented for all participants of the study, allowing a more efficient data analysis. 
\paragraph{Acknowledgements.}
This work is partially funded by the Austrian Science Fund project ``The Modeling Mind: Behavior Patterns in Process Modeling'' (P26609).
\vspace{-1em}

\bibliography{cep-web-TR}
\bibliographystyle{splncs}
\end{document}